\newif\ifarXiv
\arXivtrue   
\ifarXiv
\documentclass[12pt]{article}
\usepackage{epsfig,amsmath,amsfonts,amssymb,makeidx,ifthen,bm}
\usepackage[dvips]{color}
\usepackage{psfrag}
\usepackage[titletoc, title]{appendix}
\newtheorem{theorem}{Theorem}[section]
\newtheorem{proposition}[theorem]{Proposition}

\makeatletter
\@addtoreset{equation}{section}
\makeatother
\renewcommand{\theequation}{\thesection.\arabic{equation}}
\makeatletter
\long\def\@makecaption#1#2{{\small
\advance\leftskip1cm
\advance\rightskip1cm
\vskip\abovecaptionskip
\sbox\@tempboxa{#1: #2}%
\ifdim \wd\@tempboxa >\hsize
 #1: #2\par
\else
\global \@minipagefalse
\hb@xt@\hsize{\hfil\box\@tempboxa\hfil}%
\fi
\vskip\belowcaptionskip}}
\makeatother
\else
\documentclass[smallextended]{svjour3}  
\usepackage{graphicx,amsmath,amsfonts,amssymb,makeidx,ifthen,color}
\usepackage{psfrag,appendix,bm}
\fi
\ifarXiv
\newcommand{\jour}[2]{#1}
\else
\newcommand{\jour}[2]{#2}
\fi
\newcommand{\rmB}{\mathrm{B}}

\newcommand{\rmW}{\mathrm{W}}
\newcommand{\calG}{\mathcal{G}}
\newcommand{\calV}{\mathcal{V}}
\newcommand{\calE}{\mathcal{E}}
\newcommand{\calVB}{\calV^\rmB}

\newcommand{\calVW}{\calV^\rmW}
\newcommand{\calL}{\mathcal{L}}
\newcommand{\sfS}{\mathsf{S}}
\newcommand{\Ed}{E^\mathrm{d}}
\newcommand{\Gd}{G^\mathrm{d}}
\newcommand{\calEd}{\mathcal{E}^\mathrm{d}}
\newcommand{\calGd}{\mathcal{G}^\mathrm{d}}
\newcommand{\bcalG}{{\bar{\calG}}}
\newcommand{\bcalV}{\bar{\calV}}
\newcommand{\bcalA}{\bar{\mathcal{A}}}
\newcommand{\bcalB}{\bar{\mathcal{B}}}
\newcommand{\bcalE}{\bar{\calE}}
\newcommand{\LG}{\calL_{\bcalG}}

\newcommand{\Hhop}{H_{\mathrm{hop}}}
\newcommand{\Hint}{H_{\mathrm{int}}}
\newcommand{\up}{\uparrow}
\newcommand{\dn}{\downarrow}
\newcommand{\Ne}{{N_\mathrm{e}}}
\newcommand{\Stot}{S_\mathrm{tot}}
\newcommand{\Mtot}{M_\mathrm{tot}}
\newcommand{\bv}{{\bar{v}}}
\newcommand{\bu}{{\bar{u}}}

\newcommand{\Phiup}{\Phi_\up}
\newcommand{\Np}[2]{N_{({#1}\to {#2})}}

\newcommand{\im}{\mathrm{i}}

\newcommand{\cs}[1]{c_{#1,\sigma}}
\newcommand{\csd}[1]{c_{#1,\sigma}^\dagger}

\newcommand{\bs}[1]{b_{#1,\sigma}}
\newcommand{\bsd}[1]{b_{#1,\sigma}^\dagger}
\newcommand{\Bs}[1]{B_{#1,\sigma}}
\newcommand{\Bsd}[1]{B_{#1,\sigma}^\dagger}

\newcommand{\as}[1]{a_{#1,\sigma}}
\newcommand{\asd}[1]{a_{#1,\sigma}^\dagger}
\newcommand{\As}[1]{A_{#1,\sigma}}
\newcommand{\Asd}[1]{A_{#1,\sigma}^\dagger}
\newcommand{\tas}[1]{\tilde{a}_{#1,\sigma}}
\newcommand{\tasd}[1]{\tilde{a}_{#1,\sigma}^\dagger}
\begin{document}
\jour{ 
\noindent   
\textbf{\Large%
 An extension of the cell-construction method for the flat-band ferromagnetism 
}\bigskip\\
{Akinori Tanaka}%
\footnote
{%
Department of General Education, National Institute of Technology, 
Ariake College, Omuta, Fukuoka 836-8585, Japan\\
E-mail: akinori@ariake-nct.ac.jp
}
} 
{ 

\title{ An extension of the cell-construction method for the flat-band ferromagnetism}
\titlerunning{ An extension of the cell-construction method for the flat-band ferromagnetism}
\author{Akinori Tanaka}
%


\institute{A. Tanaka \at
              Department of General Education,
	      National Institute of Technology, Ariake College, Omuta, 
	      Fukuoka 836-8585, Japan \\
              \email{akinori@ariake-nct.ac.jp}           
}

\date{Received: date / Accepted: date}

\maketitle
} 
%
%
\begin{abstract}
We present an extension of the cell-construction method for the flat-band ferromagnetism.
In a rather general setting, 
we construct Hubbard models with highly degenerate single-electron ground states
and obtain a formal representation of these single-electron ground states.
By our version of the cell-construction method, 
various types of flat-band Hubbard models, including the one on line graphs, can be designed and
shown to have the unique ferromagnetic ground states when the electron number is equal to 
the degeneracy of the single-electron ground states. 
\jour{}
{
}
\end{abstract}
%
\jour{
\tableofcontents
\newpage
}
{}
\section{Introduction}
Rigorous results on quantum many-body systems, even if they are obtained with some special conditions, 
provide us with an understanding of mechanisms for phenomena arising from 
the interplay between the quantum mechanical motion of particles
and the interactions among them.
One of the examples is flat-band ferromagnetism found in a class of Hubbard models 
with highly degenerate single-electron ground states~\cite{Tasaki98a,Tasaki2020}.
Flat-band ferromagnetism, which was first discovered by Mielke~\cite{Mielke91a,Mielke91b,Mielke92} and Tasaki~\cite{Tasaki92}, 
explains clearly how the spin-independent Coulomb repulsion combined with the Pauli exclusion principle for electrons 
generates ferromagnetism.       
Furthermore the mechanism turns out to work in more general settings. 
In fact, it is shown that flat-band ferromagnetism is stable against perturbations which change the flat band into a dispersive band~\cite{Tasaki96,Tasaki2003,TanakaUeda2003,Lu2009,Tanaka2018,TamuraKatsura2019}.
Examples of Hubbard models which exhibit metallic ferromagnetism are also derived 
by taking into account the mechanism of flat-band 
ferromagnetism~\cite{TanakaTasaki2007,TanakaTasaki2016}. 
The idea has been further applied to more fascinating problems of 
topological systems~\cite{KatsuraMaruyamaTanakaTasaki2010,MizoguchiHatsugai2020}.

The class of flat-band ferromagnetism studied by Mielke was found 
in the flat-band Hubbard models defined on line graphs. 
The result of Mielke was obtained by using some ideas from graph theory 
and is summarized in the theorem in which the structure of graphs is
related to the occurrence of ferromagnetism in the models~\cite{Mielke91b,Mielke92}.
On the other hand, the class of Tasaki's flat-band models was constructed 
by using the cell-construction method~\cite{Tasaki92,MielkeTasaki93}.
Lattices of Tasaki's flat-band models are constructed by assembling cells 
each of which has one internal site and several external sites. 
Some external sites from different cells are identified and are regarded as 
a single site to form the whole lattice (see Fig.~\ref{fig:Tasaki}).
It is noted that every internal site belongs to exactly one cell in Tasaki's cell-construction. 
Due to this property one can obtain explicit expressions 
for the single-electron ground states,
each of which is localized around one external site 
and internal sites belonging to cells sharing the external site.
Using these expressions, 
Tasaki described clearly the mechanism which induces the ferromagnetic interaction between electrons.   
After the discovery of these concrete examples of flat-band ferromagnetism,
Mielke also developed a general theory and showed a necessary and sufficient condition for the occurrence of ferromagnetism
in general flat-band Hubbard models~\cite{Mielke93,Mielke99}.
More precisely, it was proved that the ferromagnetic ground state of the flat-band Hubbard model 
at half-filling of its flat band is the unique ground state (up to the spin degeneracy)   
if and only if the single-electron density matrix with respect to the ferromagnetic ground state is irreducible. 
This necessary and sufficient condition elegantly characterizes the occurrence of flat-band ferromagnetism.
The condition is, however, so abstract that one needs some extra work to find 
or to construct a class of flat-band Hubbard models
which satisfy this condition.  

In this paper we consider an extension of the cell-construction method for the flat-band ferromagnetism 
developed by Tasaki. 
Removing the restriction that the internal site is never shared by several cells,
which played an important role in Tasaki's cell-construction method, 
we considerably extend a class of Hubbard models which are shown to exhibit flat-band ferromagnetism.
(See Fig.~\ref{fig:example-G} and compare it with Fig.~\ref{fig:Tasaki}.)
A similar extension has been discussed in Ref.~\cite{MizoguchiHatsugai2019} to construct a class of Hubbard models with flat bands.%
\footnote{
See also Ref.~\cite{KatsuraMaruyama2015}, in which one can find a pedagogical explanation of 
how to construct the Hubbard models with flat bands.
}        
Here we concentrate on the occurrence of the unique ferromagnetic ground states as well as the construction
of flat-band Hubbard models.
By our extension, one can treat not only Tasaki's flat-band ferromagnetism but also Mielke's flat-band ferromagnetism
on line graphs in a unified manner.%
\footnote{
The present method naturally reproduces all the models in Tasaki's class.
As for the flat-band models on line graphs, on the other hand, 
there is a slight difference between the Hamiltonian in our method and that in Mielke's class.
One has to do some extra work to treat the Hamiltonian of Mielke's flat-band models 
with our cell-construction method. See also footnote~\ref{f:difference} and appendix~\ref{s:appendixA}.
}
In addition, one can construct flat-band Hubbard models which belong neither to Tasaki's class nor to Mielke's class
and can show that they have the unique ferromagnetic ground states when the electron number is equal to 
the degeneracy of the single-electron ground states. 
Although a necessary and sufficient condition for the occurrence of flat-band ferromagnetism
was already given by Mielke, our extension is useful when we consider 
a concrete example of flat-band ferromagnetism.
We hope that the present extension, 
as a complement of a general theory of Mielke, 
provides a unified viewpoint of the flat-band ferromagnetism.   

The rest of this paper is organized as follows.
In the next section we give the definition of the model and state the main results.
In section~\ref{s:Examples}, we give three examples which are shown to exhibit flat-band ferromagnetism
through our method.
In section~\ref{s:Proofs}, we prove the propositions stated in~section~\ref{s:Definition}.
In section~\ref{s:Further extensions}, we discuss possible further extensions.
In appendix~\ref{s:appendixA}, we consider flat-band ferromagnetism on line graphs through the present extension.
In appendix~\ref{s:appendixB}, in order to make the presentation self-contained, we give a proof of
the uniqueness of the ferromagnetic ground states within our notations.
\section{Definition of the model and the main result}
\label{s:Definition}
Let $G_l=(V_l,E_l)$ with $l=1,2,\dots,L$ be a complete graph,%
\footnote{
Although we use graph theoretic terminology to define a lattice (graph) 
on which our Hubbard Hamiltonian is defined, we do not use any results from graph theory and
make the paper self-contained.
} 
where $V_l$ is a set of vertices with $|V_l|\ge2$ and $E_l$ is a set of edges,%
\footnote{
The symbol $|X|$ denotes the number of elements in $X$.
}
unordered pairs of two vertices, given by
\begin{equation}
 E_l=\{e=\{v,v^\prime \}~|~v,v^\prime \in V_l, v\ne v^\prime \}. 
\end{equation}
By using complete graphs $G_1,G_2,\dots,G_L$, we construct a sequence of 
coloured graphs $\calG_1, \calG_2, \dots, \calG_L$
whose vertices are painted with either black or white.
To start with, we choose a vertex, which we label as $v_l^0$, from each $V_l$. 
We then paint the vertices $\{v_l^0\}_{l=1}^L$ black and paint all the other vertices white.
\begin{figure}
 \begin{center}
  \includegraphics[width=.9\textwidth]{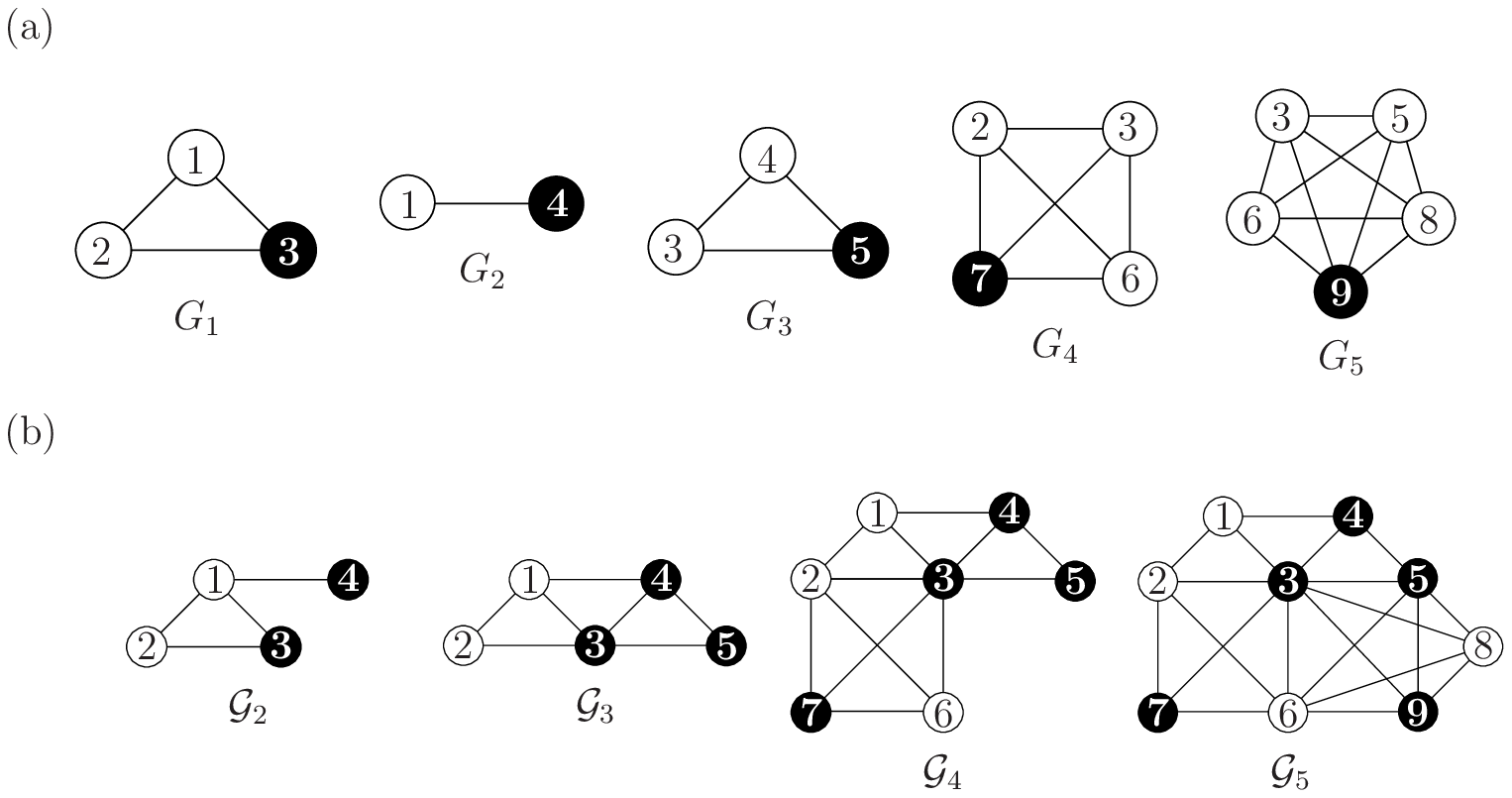}
 \end{center}
\caption{
An example with 5 complete graphs. Circles and lines represent vertices and edges, respectively.
(a) One vertex $v_l^0$ in each complete graph $G_l$ is painted with black. In this example, $v_1^0=3,v_2^0=4,v_3^0=5,v_4^0=7$, and $v_5^0=9$. Identifying vertices that are labeled with the same number, we inductively construct the sequence of the coloured graphs.
(b) Coloured graphs $\calG_2,\dots,\calG_5$ made of $G_1,\dots,G_5$ in (a).
The weights $w(v)$ of vertices are one for $v=7,8,9$, two for $v=1,2,4,5,6$, and four for $v=3$. 
The weights $w(e)$ of edges are two for $e=\{2,3\},\{3,5\},\{3,6\}$, and one for other edges.}
\label{fig:example-G}
\end{figure}

We construct coloured graphs $\left\{\calG_l\right\}_{l=1}^L$ by an inductive way. The vertex set and the edge set corresponding to $\calG_l$ 
are denoted by $\calV_l$ and $\calE_l$, respectively. 
To represent a colour of each vertex $v\in\calV_l$ we introduce maps $\gamma_l$ from $\calV_l$ to 
the colour set $\{\mathrm{B(black),W(white)}\}$ for $l=1,2,\dots,L$.
The colour map $\gamma_l$ will be explicitly defined in each step of the construction of $\calG_l$. 
First,  we set $\calG_1=(V_1,E_1,\gamma_1)$ with the colour map
\begin{equation}
 \gamma_1(v)=\left\{
	      \begin{array}{ll}
	       \rmB & \mbox{if $v=v_1^0$},\\
	       \rmW & \mbox{otherwise}.
	      \end{array}
		\right.
\end{equation}
Then, for $l=2,\dots,L$, from the so far constructed coloured graph 
$\calG_{l-1}=(\calV_{l-1},\calE_{l-1},\gamma_{l-1})$ and the complete graph $G_{l}=(V_l,E_l)$, 
we generate $\calG_{l}=(\calV_l,\calE_l,\gamma_l)$ in the following manner.   
Let $z_l$ be an integer with $0<z_l\le |V_l|-1$ 
and choose $z_l$ white vertices in $V_l$ of the complete graph $G_l$.
We identify each of the chosen white vertices with an arbitrary vertex (which may be black or white) in $\calV_{l-1}$ 
and regard it as a single vertex in $\calV_l$.
The vertex set $\calV_l$ is defined as the collection of vertices in $\calV_{l-1}$ and $V_l$ with the above identification.   
We note that $\calV_l=\calV_{l-1}\cup V_l=\cup_{m=1}^{l}V_m$ and that $|\calV_l|=|V_1|+\sum_{m=2}^l(|V_m|-z_m)$.
The edge set $\calE_l$ is defined by $\calE_l=\calE_{l-1}\cup {E_l}=\cup_{m=1}^l E_m$ 
where two edges touching the same two vertices (as a result of the above vertex identification) are merged into a single edge.%
\footnote{Namely, we do not consider multiple edges in $\calG_l$. Instead, we later introduce weights of edges. 
}
The colour map $\gamma_l$ of $\calG_l$ is defined by
\begin{equation}
 \gamma_l(v) = \left\{
		\begin{array}{ll}		
		 \gamma_{l-1}(v) & \mbox{if $v\in\calV_{l-1}$,}\\
		 \rmB & \mbox{if $v=v_{l}^0$,} \\
		 \rmW & \mbox{otherwise.}
		  \end{array}
 \right.
\end{equation}
In other words, a white vertex and a black vertex are merged into a black one,  
while two white vertices are merged into a white one, in the vertex identification.
We write $\calVB_l$ for the set of black vertices in $\calV_l$ and similarly write $\calVW_l$ 
for the set of white vertices. 
Note that the black vertex $v_m^0$ of $G_m$ has not been merged  
with vertices in $\calV_{m-1}$ when constructing $\calG_m$ for all $m=2,\dots,l$.
We thus use $\{v_m^0\}_{m=1}^l$ as a label for vertices in $\calVB_l$.
For $v\in\calV_L$, let $w(v)$ denote the number of subsets $V_l$ which contain the vertex $v$.%
\footnote{%
Here and in what follows, $V_l$ and  $E_l$ with $l=1,2,\dots,L$ are used to represent subsets of $\calV_L$ and $\calE_L$, respectively. 
The elements of $V_l$ and $E_l$, when we regard them as subsets, are vertices and edges that come from the complete graph $G_l=(V_l,E_l)$. 
}
We call $w(v)$ a weight of the vertex $v$. 
Similarly, for $e\in\calE_L$ we denote by $w(e)$ the number of subsets $E_l$ which contain the edge $e$ and call it
a weight of the edge~$e$. 
See Fig.~\ref{fig:example-G} for an example.

We consider the Hubbard model on the graph $\calG_L$.
Let $c_{v,\sigma}$ and $c_{v,\sigma}^\dagger$ be an annihilation operator 
and a creation operator of an electron with spin $\sigma=\up,\dn$ at the vertex $v\in\calV_L$.
They are assumed to satisfy the usual anticommutation relations
\begin{equation}
\{c_{v,\sigma},c_{v^\prime,\tau}\}=\{c_{v,\sigma}^\dagger,c_{v^\prime,\tau}^\dagger\}=0   
\end{equation}
and
\begin{equation}
\label{eq:c-anticommutation}
\{c_{v,\sigma}^\dagger,c_{v^\prime,\tau}\}=\delta_{v,v^\prime}\delta_{\sigma,\tau}
\end{equation}
for $v,v^\prime\in \calV_L$ and $\sigma,\tau=\up,\dn$.
The operator $n_{v,\sigma}=c_{v,\sigma}^\dagger c_{v,\sigma}$ 
is the number operator of an electron with spin $\sigma$ at $v$.
We denote by $\Ne$ the total number of electrons on $\calV_L$ and denote by $\Phi_0$ a state with no electrons.
The spin operators at a vertex $v$ are defined by 
$S_v^{(1)}=(S_v^+ + S_v^-)/2, S_v^{(2)}=(S_v^+ - S_v^-)/(2\im), S_v^{3}=(n_{v,\up}-n_{v,\dn})/2$ with
$S_v^+=c_{v,\up}^\dagger c_{v,\dn},S_v^-=c_{v,\dn}^\dagger c_{v,\up}$.
The total spin operators are defined by $\Stot^{(\alpha)}=\sum_{v\in\calV_L} S_{v}^{(\alpha)}$ with $\alpha=1,2,3$.
The eigenvalues of $\Stot^{(3)}$ and $(\Stot^{(1)})^2+(\Stot^{(2)})^2+(\Stot^{(3)})^2$ are 
denoted by $\Mtot$ and $\Stot(\Stot+1)$, respectively.
 
The Hubbard Hamiltonian we consider is $H=\Hhop+\Hint$ with
\begin{equation}
\Hhop=\sum_{v,v^\prime\in\calV_L}\sum_{\sigma=\up,\dn} t_{v,v^\prime}\csd{v}\cs{v^\prime}
\end{equation}
and
\begin{equation}
\Hint=U \sum_{v\in\calV_L}n_{v,\up}n_{v,\dn},
\end{equation}
where $U>0$ and hopping amplitudes $t_{v,v^\prime}$ are given by
\begin{equation}
 t_{v,v^\prime}=\left\{
	  \begin{array}{@{\,}ll}
	   w(e)t & \mbox{if $e=\{v,v^\prime\}\in\calE_L$,}\\
           w(v)t       & \mbox{if $v=v^\prime\in\calV_L$,}\\
           0 & \mbox{otherwise,}
	  \end{array}
	 \right.
\end{equation}
with positive parameter $t$.
For this Hubbard Hamiltonian we have the following propositions.
\begin{proposition}
\label{proposition1}
 The lowest single-electron energy of the hopping Hamiltonian $\Hhop$ is zero 
 and is $(|\calV_L|-L)$-fold degenerate.  
\end{proposition}
\begin{proposition}
\label{proposition2}
Consider the Hamiltonian $H=\Hhop+\Hint$ with $\Ne=|\calV_L|-L$. 
A ferromagnetic state with $\Mtot=\Ne/2$ where every single-electron ground state of $\Hhop$  
is singly occupied by an $\up$-spin electron is one of the ground states.   
\end{proposition} 

We note that the above proposition does not claim the uniqueness of the ground state.
In fact, we need a certain condition on the graphs $\{\calG_l\}_{l=1}^L$ to show 
the uniqueness of the ground state.

Let us introduce some more notations to state the condition 
which guarantees the uniqueness of the ferromagnetic ground states.

Let $(v,v^\prime)$ be an ordered pair of vertices, a directed edge, and define $\Gd_l=(V_l,\Ed_l)$ with
\begin{equation}
 \Ed_l=\{(v,v_l^0)~|~v\in V_l\backslash\{v_l^0\}\}
\end{equation}
for $l=1,\dots,L$. Every directed edge in $\Gd_l$ is pointing from a white vertex to the black vertex $v_l^0$.
Then, in the same way as that used for $\calG_l$, we inductively construct directed coloured graphs $\calGd_l=(\calV_l,\calEd_l,\gamma_l)$, 
replacing $\{\calE_m\}_{m=1}^l$ and $\{E_m\}_{m=1}^l$ with
$\{\calEd_m\}_{m=1}^l$ and $\{\Ed_m\}_{m=1}^l$, respectively. 
We note that it is not necessary to merge two directed edges in the construction process of $\calGd_l$ 
since there are no directed edges connecting two white vertices in $\Ed_m$.

Now we consider the directed coloured graph $\calGd_L$ corresponding to $\calG_L$ on which our Hubbard model is defined. 
We say that a vertex $u$ is reachable from a vertex $v$ if there exists a sequence of vertices, called a directed path, 
$(v\to u)=(v_0,v_1,\dots,v_{k-1},v_k)$, such that $v_0=v,v_k=u$ and $(v_{j-1},v_j)\in\calEd_L$ for all $j=1,\dots,k$.  
For each $v\in\calV_L^\rmW$ we denote by $R(v)$ the set of vertices which are reachable from $v$.
For later convenience, $v$ is included in $R(v)$ as its element with a path $(v\to v)=(v)$. 
It is noted that a colour of every vertex in $R(v)\backslash\{v\}$ is black. 
It is also noted that, 
for $u$ in $R(v)$, there might be several directed paths from $v$ to $u$. 
We denote by $\Np{v}{u}$ the number of directed paths from $v\in\calVW_L$ to $u\in R(v)$.
\begin{figure}
 \begin{center}
  \includegraphics[width=.4\textwidth]{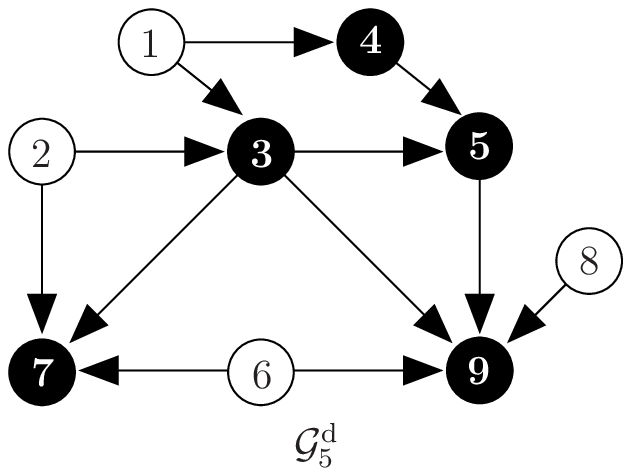}
 \end{center}
\caption{
The directed coloured graph $\calGd_5$ corresponding to $\calG_5$ in Fig.~\ref{fig:example-G}(b). 
For this example, we have $R(1)=\{3,4,5,7,9\}, R(2)=\{3,5,7,9\}, R(6)=\{7,9\}$ and $R(8)=\{9\}$.
Noting that $z_l<|V_l|-1$ for $l=4,5$, one finds that the assumption (A1) is satisfied,
since the white vertex $1\in\calVW_3$ is reachable to $v_4^0=7$ by the single path $(1\to 7)=(1,3,7)$ 
and the white vertex $6\in\calVW_4$ is reachable to $v_5^0=9$ by the single path $(6\to9)=(6,9)$.
(The white vertex $1\in\calVW_4$ is also reachable to $v_5^0$ by three paths.) 
It thus follows from Proposition~\ref{proposition3} that the Hubbard Hamiltonian $H$ with $\Ne=4$ 
has the unique ferromagnetic ground states. 
}
\label{fig:example-Gd}
\end{figure}

We suppose that the coloured directed graph $\calGd_L$ possesses the following property:
\bigskip\\\noindent
(A1) For every $l\in \{2,\dots,L\}$ such that $z_l\ne|V_l|-1$, 
there exists a white vertex $v\in\calV_{l-1}^\rmW$ 
from which the black vertex $v_l^0\in\calVB_L$ 
is reachable by an odd number $\Np{v}{v_l^0}$ of directed paths.
\bigskip\\\noindent
See Fig.~\ref{fig:example-Gd} for an example.  
It is remarked that $\Np{v}{v_l^0}$ is determined and fixed when we construct $\calGd_l$.%
\footnote{
The set of directed edges connecting two vertices in $\calV_l$ does not change 
in the remaining process of construction, since
directed edges in $\calEd_L\backslash \calEd_l$ are written as $(u,v_m^0)$ with $m=l+1,\dots,L$
whose end-points are always in the outside of $\calV_l$.
}
Thus we can check whether the assumption (A1) is satisfied or not for the newly added black vertex 
$v_l^0$ at each step of the construction of the graph $\calGd_l$.
Then, we have the following proposition.
\begin{proposition}
\label{proposition3}
Under the assumption (A1) 
the ground state of the Hamiltonian $H=\Hhop+\Hint$ with $\Ne=|\calV_L|-L$ 
has $\Stot=\Ne/2$ and is unique apart from the degeneracy due to the spin-rotation symmetry.   
\end{proposition} 
The occurrence of flat-band ferromagnetism in the present model is roughly explained as follows.
The highly degenerate single-electron ground states of flat-band models are known to be localized.
In the present model, each of them is localized around a white vertex, 
as we will see later in section~\ref{s:Proofs}.
In order to avoid the increase of the interaction energy due to the double occupancy of white vertices, 
each electron singly occupies a localized state.
Then, two electrons occupying localized states which overlap with each other at some black vertex 
align their spins in order to avoid the energy increase due to the double occupancy of that black vertex.
Now we say that two localized states are directly connected if they overlap with each other.
The assumption (A1) guarantees that 
any two single-electron ground states are connected in the above sense, 
and thus whole electrons form ferromagnetic ground states.
\section{Examples}
\label{s:Examples}
\begin{figure}
 \begin{center}
  \includegraphics[width=.9\textwidth]{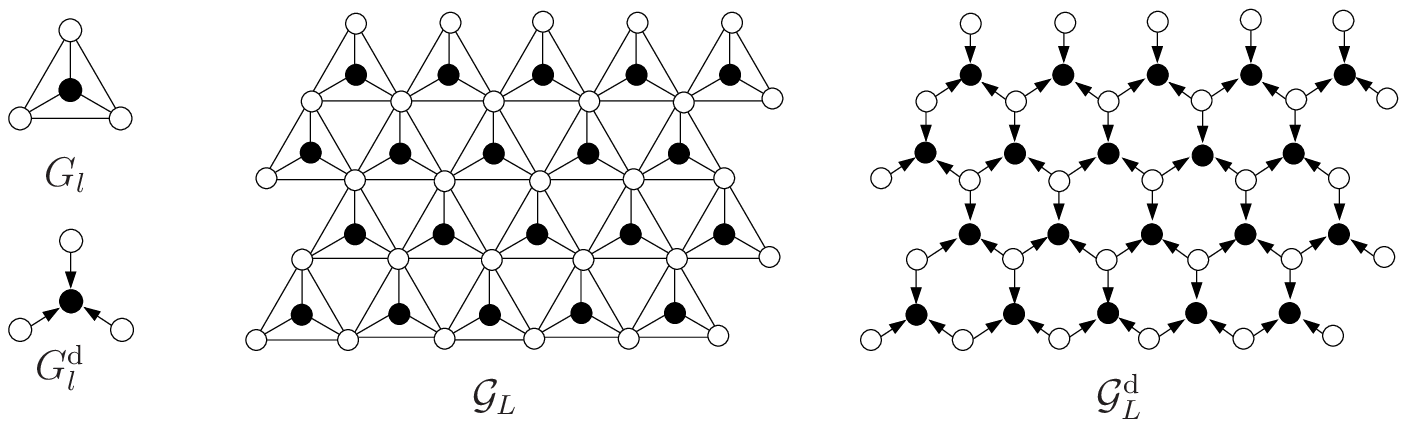}
 \end{center}
\caption{
An example of the lattice structure in Tasaki's flat-band models. 
The lattice is constructed by identifying only white vertices (external sites) of the 4-vertex complete graphs (4-site cells). 
}
\label{fig:Tasaki}
\end{figure}

\textit{Tasaki's flat-band models.}
The main idea of our constructive method comes from 
Tasaki's flat-band models~\cite{Tasaki92,MielkeTasaki93}, which are, of course, possible to construct
following our procedure and shown to satisfy the assumption (A1).%
\footnote{
To be more precise, $\Hhop^\prime(\sfS) + \Hint$ where $\Hhop^\prime(\sfS)$ is given in \eqref{eq:Hhopprime}
with $\sfS_{l,l^\prime}=t\delta_{l,l^\prime}$
corresponds to the Hamiltonian of Tasaki's flat-band models.
} 
Complete graphs $G_l$, black vertices $v_l^0$ and white vertices 
correspond to cells, internal sites and external sites, respectively, in Tasaki's flat-band models.
A coloured graph corresponding to a lattice of Tasaki's flat-band models 
is constructed by identifying only white vertices (external sites) 
from different complete graphs (cells) to regard them as a single vertex (site) in the coloured graph. 
In Fig.~\ref{fig:Tasaki}
we show an example in which the coloured graph is constructed by 4-vertex complete graphs (4-site cells).      
We note that, since black vertices are never identified with other vertices in this case, 
one can construct a coloured graph $\calG_L$ of any shape with any boundary conditions, 
without worrying about how to add a complete graph $G_l$ to the so far constructed coloured graph $\calG_{l-1}$. 
It is also easy to check that the assumption (A1) holds for the coloured directed graph $\calGd_L$ 
corresponding to $\calG_L$ constructed with the above restriction. 
In fact, we have 
$R(v)=\{u~|~\mbox{$u=v$ or $(v,u)\in\calEd_L$}\}$, which implies $\Np{v}{u}=1$, for all $v\in\calVW_L$ in this case
and thus there exists $v\in\calVW_{l-1}$ such that $\Np{v}{v_l^0}=1$ for all $v_l^0\in\calVB_L$. 
As was already proved, Proposition~\ref{proposition3} also shows that 
Tasaki's flat-band models have the unique ferromagnetic ground states if the electron number $\Ne$ is equal to 
the degeneracy of the single-electron ground states, i.e., the number of
white vertices, $|\calV_L|-L$, in $\calG_L$. 

\textit{Kagom\'e lattice with open boundary conditions.}
Our constructive method can treat models corresponding 
not only to Tasaki's flat-band models but also to Mielke's flat-band models, the flat-band models on line graphs.%
\footnote{
Every line graph is characterized by its complete subgraphs, with which we can associate $\{G_l\}_{l=1}^L$.
Our method is applicable when the line graph can be reconstructed from $\{G_l\}_{l=1}^L$ following
our procedure stated in section~\ref{s:Definition}. See also appendix~\ref{s:LineGraph}.
}
A typical example of line graphs is the kagom\'e lattice and the Hubbard model on it
was studied by Mielke in detail.
We here consider the Hubbard model on the kagom\'e lattice with open boundary conditions.%
\footnote{
Modifying our method further, we can treat the periodic case as well. See appendix~\ref{s:appendixA}.
}
In Fig.~\ref{fig:kagome}(a), we show the kagom\'{e} lattice, which is the line graph of a hexagonal lattice $G_\mathrm{hex}$,
with the boundary. 
We note that one can not directly apply the theorem of Mielke in Refs.~\cite{Mielke91b} and \cite{Mielke92} 
to conclude the uniqueness of the ferromagnetic ground states since the hexagonal lattice $G_\mathrm{hex}$ 
which we use in this example is not 2-connected due to the vertices in the boundary,
each of which is connected to only one other vertex in the inside.   

Let us treat 
the flat-band Hubbard model on the line graph of $G_\mathrm{hex}$
through our method. 
Let $l_1$ and $l_2$ be positive integers. 
We construct the coloured graph $\calG_L$ with $L=2l_1\times l_2$ 
depicted in Fig.~\ref{fig:kagome}~(b) in the following manner.
We first make the bottom part $\calG_{2l_1}$, 
which is a chain of triangles.
To do so, placing the black vertex $v_l^0$ in $V_l$ at the right most position, 
we identify the right most black vertex $v_{l-1}^0$ in $\calV_{l-1}$ 
with a white vertex in $V_l$ for all $l=2,\dots,2l_1$.  
We note that $z_l=1$ for $l=2,\dots,2l_1$. 
Next, we add $G_{2l_1+1}$ to the second left most triangle in $\calG_{2l_1}$ by identifying a white vertex in $V_2$
with a white one in $V_{2l_1+1}$.  
Then, for $l=2l_1+2,\dots,4l_1$, if $l$ is even, we identify $v_{l-1}^0$ with a white vertex in $V_l$ 
and, if $l$ is odd, we identify not only $v_{l-1}^0$ with one white vertex in $V_l$ but also the white vertex in $V_{l-2l_1+1}$ 
in the bottom part with the other white vertex in $V_l$.
We note that $z_{2l_1+1}=1$ and, for $l=2l_1+2,\dots,4l_1$, $z_l=1$ if $l$ is even and $z_l=2$ otherwise.
Repeating a similar procedure to that in making $\calG_{4l_1}$, we obtain $\calG_L$ with $L=2l_1\times l_2$.    
It is easy to see from Fig.~\ref{fig:kagome}~(b) that the assumption (A1) is satisfied for $\calGd_L$.
Therefore, we can conclude from Proposition~\ref{proposition3} that 
the Hubbard model $H=\Hhop+\Hint$ on the kagom\'e lattice with the open boundary conditions 
exhibits the flat-band ferromagnetism.%
\footnote{
Strictly speaking, there is a minor difference between $H$ and the Hamiltonian of Mielke's flat-band model;
the on-site potential at the boundary is different from the inside one in $H$ 
while the on-site potential is uniform in Mielke's flat-band model. 
With some extra work we can also show that $H+t\sum_{v\in\calV_L;w(v)=1}\sum_{\sigma}\csd{v}\cs{v}$, 
in which the on-site potential is uniform, 
has the unique ferromagnetic ground states if $\Ne$ is equal to the number of hexagons
in $\calG_L$.
\label{f:difference}}  
\begin{figure}
\begin{center}
 \includegraphics[width=.9\textwidth]{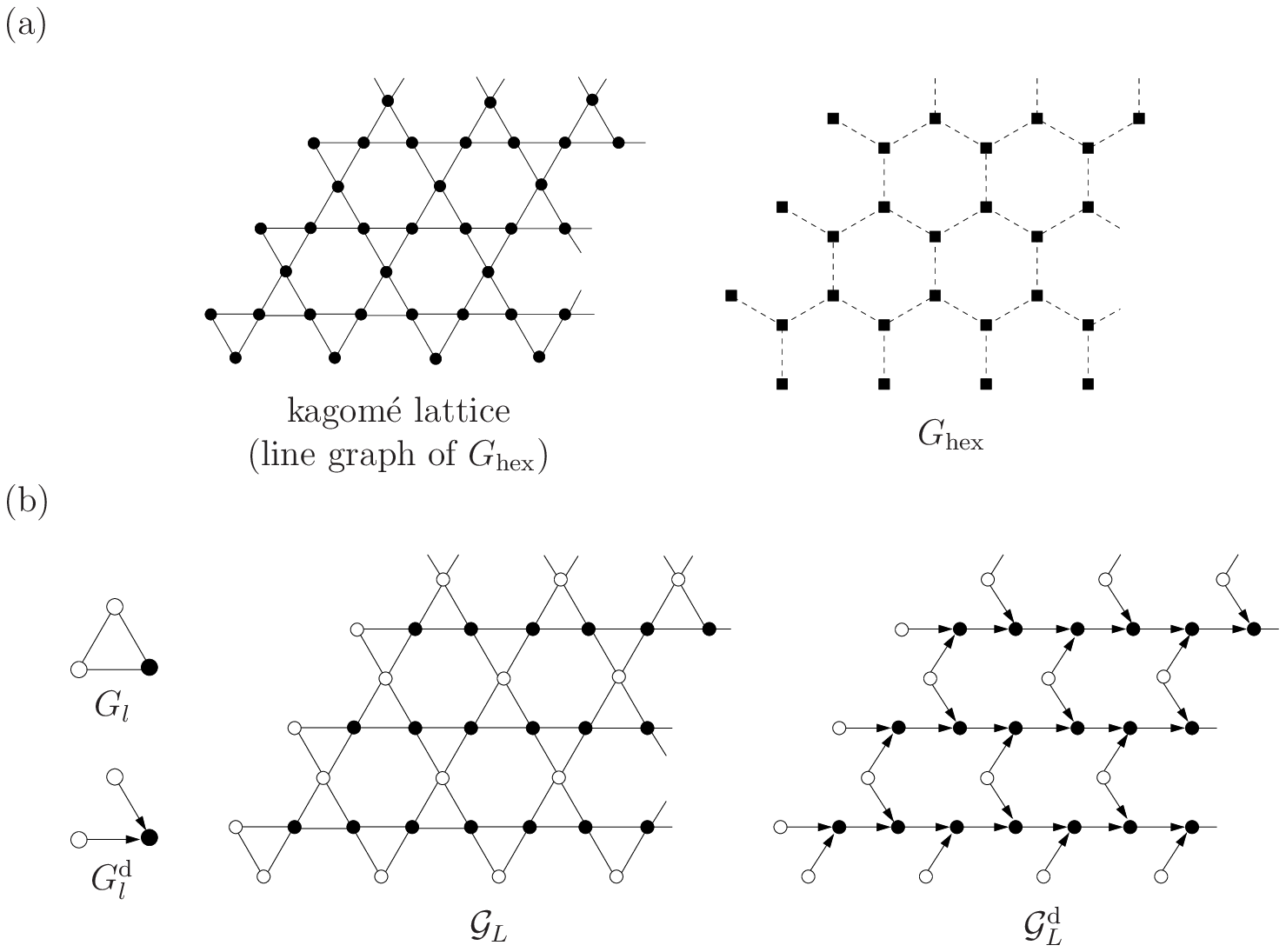}
\end{center}
\caption{
 (a) The kagom\'e lattice with the boundary is obtained by associating a vertex (solid circle) 
 with each edge (dashed line) of $G_\mathrm{hex}$
 and connecting two vertices (solid circles) with an edge (solid line) 
 if the corresponding edges (dashed lines) of $G_\mathrm{hex}$ have a vertex (solid square) in common. 
 The boundary edges of $G_\mathrm{hex}$ are arranged 
 so that the line graph of $G_\mathrm{hex}$ may coincide with the structure of $\calG_L$ in Fig.~\ref{fig:kagome}(b).
 (b) 
 By using 3-vertex complete graphs, one can construct the coloured graph $\calG_L$ whose lattice structure is 
 the kagom\'e lattice with the boundary. 
}
\label{fig:kagome}
\end{figure}

\begin{figure}
 \begin{center}
  \includegraphics[width=.9\textwidth]{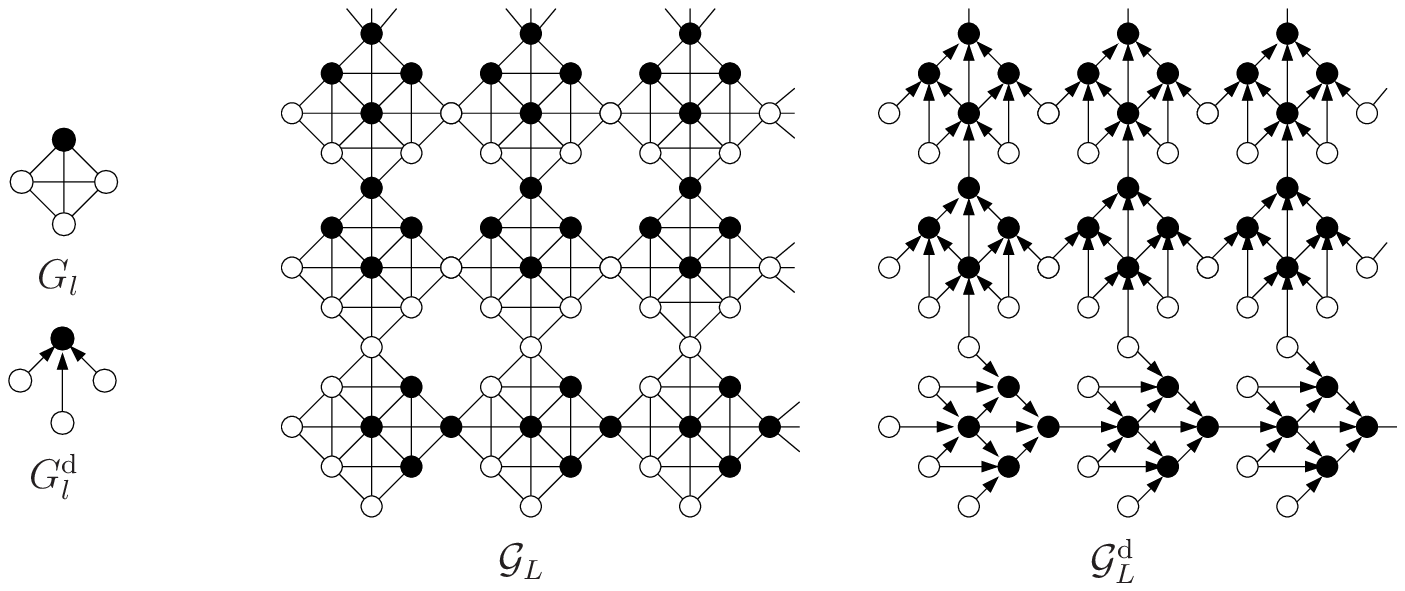}
 \end{center}
 \caption{
 A variant of the checkerboard lattice which is composed of 4-vertex complete graphs.
}
\label{fig:K4}
\end{figure}

\textit{A variant of the checkerboard lattice composed of 4-vertex complete graphs.}
Finally, we present an example which is classified as neither Tasaki's nor Mielke's flat-band models.
The lattice structure is depicted in Fig.~\ref{fig:K4}.
As we can see, 
$\calG_L$ is not an assembly of cells with internal sites, which play a crucial role in Tasaki's models, and
it is also impossible to regard $\calG_L$ as a line graph of any graph.
The coloured graph $\calG_L$ is constructed in the following manner. 
We first make the bottom part of $\calG_L$ by adding the 4-vertex complete graphs to the so far constructed coloured graph as
\begin{center}
 \includegraphics[width=.8\textwidth]{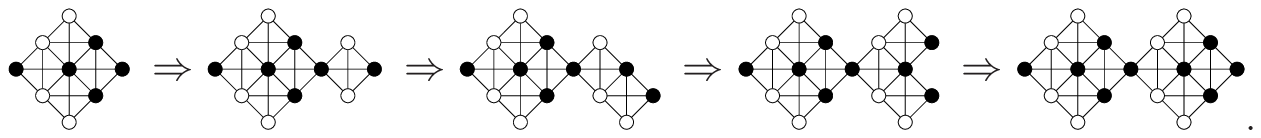}.
\end{center}
Then, to this bottom part, we add 4-vertex complete graphs as
\begin{center}
 \includegraphics[width=.85\textwidth]{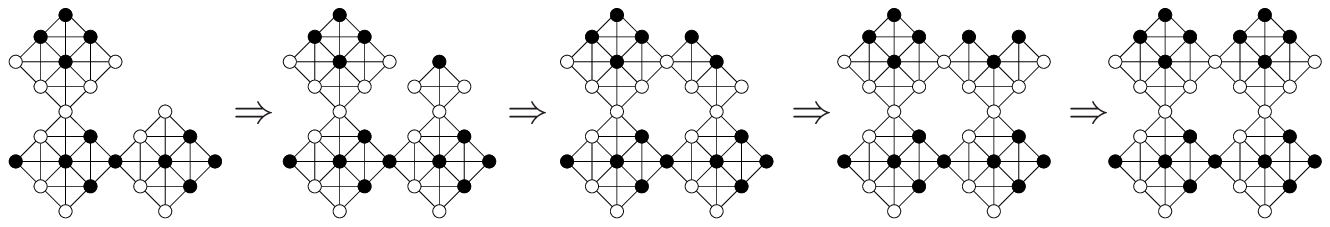}
\end{center}
in order to make the second part from the bottom.
Repeating a similar procedure to that in making the second part, we get the coloured graph $\calG_L$. 
We find from Fig.~\ref{fig:K4} that $\calGd_L$ corresponding $\calG_L$ satisfies (A1), 
and thus the Hubbard model $H=\Hhop+\Hint$ with $\Ne=|\calV_L|-L$ has the unique ferromagnetic ground states.   
\section{Proofs}
\label{s:Proofs}
\textit{Proof of Proposition~\ref{proposition1}.}
For each subset $V_l$ of $\calV_L$, let us define a fermion operator by
\begin{equation}
 \label{eq:b-operator}
 \bs{l} = \sum_{v\in V_l} \cs{v}.
\end{equation}
By using the $b$ operators, we can rewrite $\Hhop$ as%
\footnote{
In fact we set the hopping amplitudes of our Hubbard Hamiltonian so that we can rewrite $\Hhop$ 
into the form of \eqref{eq:Hhop}.}
\begin{equation}
\label{eq:Hhop}
 \Hhop=t\sum_{l=1}^L \sum_{\sigma=\up,\dn} \bsd{l}\bs{l}. 
\end{equation}
From this expression of $\Hhop$, one finds that $\Hhop$ is a positive semidefinite operator. 

Let us show that $\{\bs{l}\}_{l=1}^L$ is linearly independent, 
i.e., the equation $\sum_{l=1}^L \beta_l \bs{l}=0$ holds if and only if $\beta_l=0$ for all $l=1,\dots,L$.
To see this, first note that we have $\{c_{v_{m}^0,\sigma}^\dagger, \bs{l}\}=\delta_{m,l}$ for $1\le l\le m$, 
since the black vertex $v_{m}^0$ is not contained in $\calV_{l}$ with $1\le l < m$.
Then, assuming $\sum_{l=1}^L \beta_l \bs{l}=0$, we have $\beta_L=0$ from $\{c_{v_L^0,\sigma}^\dagger, \sum_{l=1}^L \beta_l \bs{l} \}=0$, 
$\beta_{L-1}=0$ from $\{c_{v_{L-1}^0,\sigma}^\dagger, \sum_{l=1}^{L-1} \beta_l \bs{l} \}=0$, and so forth.
Therefore, $\sum_{l=1}^L \beta_l \bs{l}=0$ implies $\beta_l=0$ for all $l=1,\dots,L$.
Since the converse is trivial, we conclude that $\{\bs{l}\}_{l=1}^L$ is linearly independent.

Since the dimension of the single-electron Hilbert space on $\calV_L$ is $|\calV_L|$, 
we can find $|\calV_L|-L$ fermion operators which anticommute with all the $b$ operators.
These operators correspond to single-electron states having the lowest energy zero of the positive semidefinite operator $\Hhop$.
This completes the proof of Proposition~\ref{proposition1}.
\bigskip

Before proceeding to the proofs of Propositions \ref{proposition2} and \ref{proposition3}, 
let us consider explicit forms of the fermion operators which anticommute with the $b$ operators.
For $v\in\calV_L^\rmW$ and $u\in R(v)$, let us denoted by $(v\to u)_{j}$ with $j=1,\dots, \Np{v}{u}$ a path 
by which $u$ is reachable from $v$.
We use $|(v\to u)_{j}|$ to denote the number of vertices constituting a path $(v\to u)_{j}$.  
Then, for each $v\in\calV_L^\rmW$, we define
\begin{equation}
\label{eq:a-operator}
 \as{v}= \sum_{u\in R(v)}\left(\sum_{j=1}^{\Np{v}{u}}(-1)^{|(v\to u)_j|-1}\right) \cs{u}.
\end{equation}
\begin{figure}
 \begin{center}
  \includegraphics[width=.9\textwidth]{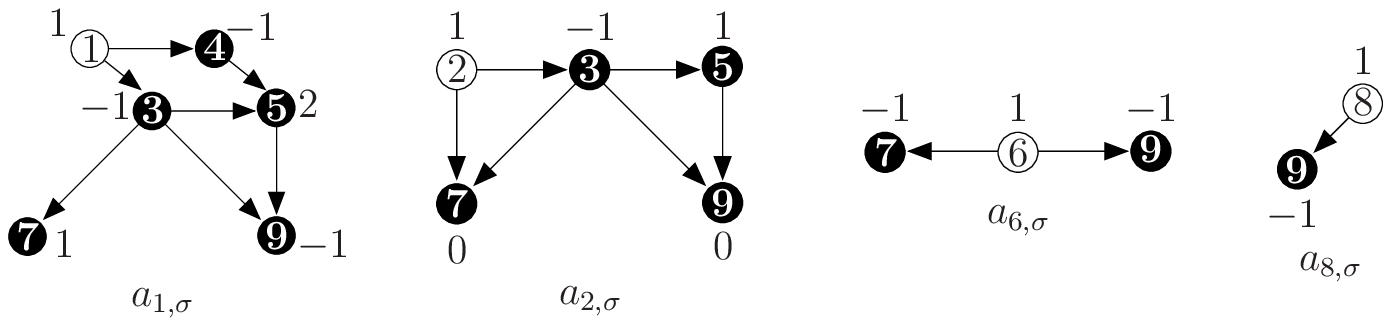}
 \end{center}
\caption{$a$ operators corresponding to $\calGd_5$ in Fig.~\ref{fig:example-Gd}.
The number shown near the vertex $u$ is the coefficient $\sum_{j=1}^{\Np{v}{u}}(-1)^{|(v\to u)_j|-1}$
of $\cs{u}$ in $\as{v}$. 
}
\label{fig:a operators}
\end{figure}
See Fig.~\ref{fig:a operators} for $a$ operators corresponding to $\calGd_5$ in Fig.~\ref{fig:example-Gd}. 
Let us show that the $a$ operators defined above anticommute with $b$ operators.
By a direct calculation with \eqref{eq:c-anticommutation} we have
\begin{equation}
 \label{eq:ab-anticommutation}
 \{\asd{v},\bs{l}\}=\sum_{u\in R(v)\cap V_l}
 \left(\sum_{j=1}^{\Np{v}{u}}(-1)^{|(v\to u)_j|-1}\right).
\end{equation}
The right hand side of \eqref{eq:ab-anticommutation} is apparently zero when $R(v)\cap V_l=\emptyset$.  
Now suppose that $R(v)\cap V_l\ne\emptyset$. In this case, $v_l^0$ is inevitably contained in $R(v)\cap V_l$, 
since, by definition, $v_l^0$ is reachable from every other vertex in the subset $V_l$. On the other hand, 
$R(v)\cap V_l$ contains at least one vertex except for $v_l^0$, since a directed path from $v$ 
must go through another vertex in the subset $V_l$ to reach $v_l^0$.  
This also means that a directed path from $v$ to $v_l^0$ is always written as 
\begin{equation}
(v\to v_l^0)=(v,\dots,u,v_l^0)
\end{equation}
with a certain vertex $u$ in $(R(v)\cap V_l)\backslash\{v_l^0\}$.
Therefore, by using $|(v,\dots,u,v_l^0)|=|(v,\dots,u)|+1$,%
\footnote{
When $v=u\in V_l$, we have $(v\to v_l^0)=(v,v_l^0)$ and $|(v\to v_l^0)|=|(v)|+1=2$.
}
the right hand side of \eqref{eq:ab-anticommutation} becomes
\begin{eqnarray}
&&
 \sum_{u\in (R(v)\cap V_l)\backslash\{v_l^0\}}
 \left(\sum_{j=1}^{\Np{v}{u}}(-1)^{|(v\to u)_j|-1}\right)
 + 
 \left(\sum_{j=1}^{\Np{v}{v_l^0}}(-1)^{|(v\to v_l^0)_j|-1}\right)
\nonumber\\ 
&&=
 \sum_{u\in (R(v)\cap V_l)\backslash\{v_l^0\}} 
  \left(
   \sum_{j=1}^{\Np{v}{u}}
   \left(
    (-1)^{|(v\to u)_j|-1}
    + (-1)^{(|(v\to u)_j|+1)-1}
   \right)
  \right)=0
\end{eqnarray}
which proves the desired claim that $\{\asd{v},\bs{l}\}=0$ for all $l=1,\dots,L$.

Noting that $\{\csd{u},\as{v}\}=\delta_{u,v}$ for any $u,v\in\calV_L^\rmW$, 
we find that $\{\as{v}\}_{v\in\calV_L^\rmW}$ is linearly independent.
Since $L=|\calVB_L|$ by our construction method of $\calG_L=(\calV_L,\calE_L,\gamma_L)$, we have ${|\calV_L|-L=|\calVW_L|}$.
This means that $\{\as{v}\}_{v\in\calVW_L}$ forms a complete set of the fermion operators corresponding to 
the single-electron zero-energy states of $\Hhop$. 
\bigskip

\textit{Proof of Proposition~\ref{proposition2}.}
Assume that $\Ne$ is fixed to $|\calV_L|-L$ and define 
\begin{equation}
 \Phiup = \left(\prod_{v\in\calVW_L} a_{v,\up}^\dagger\right) \Phi_0.
 \label{eq:ferro}
\end{equation}
Since $\Hhop$ and $\Hint$ are positive semidefinite, a zero-energy state for both of these operators, if it exists, is a ground state
of $H=\Hhop+\Hint$.
It is easy to see that $\Phiup$ is indeed a zero-energy state for both $\Hhop$ and $\Hint$. This completes the proof of Proposition~\ref{proposition2}.
\bigskip

\textit{Proof of Proposition~\ref{proposition3}.}
Under the assumption (A1) we shall show that the $a$ operators $\{\as{v}\}_{v\in\calVW_L}$ 
are connected in the following sense.
We say that $\as{v}$ and $\as{v^\prime}$ are directly connected at $u$ if there exists a vertex $u$ such that 
they have non-vanishing coefficients $\{\csd{u},\as{v}\}$ and $\{\csd{u},\as{v^\prime}\}$ of $\cs{u}$ 
in their expression~\eqref{eq:a-operator}.
We note that the colour of a vertex $u$ at which $a$ operators are connected is always black.    
The $a$ operators $\{\as{v}\}_{v\in\calV^\prime}$, where $\calV^\prime$ is a subset of $\calVW_L$, 
are said to be connected if, for any $v,v^\prime\in\calV^\prime$, there exists a sequence of 
vertices $v_0,v_1,\cdots,v_k$ such that $v_0=v,v_k=v^\prime$, and $\as{v_{j-1}}$ and $\as{v_{j}}$ are directly connected for all $j=1,\dots,k$.   
From the result of Mielke,%
\footnote{
See also Ref.~\cite{Tasaki2020} for pedagogical explanations. 
We also give a proof with our notations in appendix~\ref{s:appendixB}.
}
the connectivity of $a$ operators $\{\as{v}\}_{v\in\calVW_L}$ 
implies the uniqueness of the ferromagnetic ground state~\cite{Mielke93}, 
and thus our proposition shall be proved. 

Recall that the sequence $\calVW_1, \calVW_2,\dots,\calVW_L$ is increasing, i.e., 
$\calVW_1 \subseteq \calVW_2\subseteq\cdots\subseteq\calVW_L$.
Let us show that the connectivity of the $a$ operators $\{\as{v}\}_{v\in\calVW_{l-1}}$ implies that of $\{\as{v}\}_{v\in\calVW_{l}}$.   
In the case $\calVW_{l-1}=\calVW_{l}$, where $z_l=|V_l|-1$, the above claim is trivial.
Now suppose that $v^\prime\in\calVW_{l}\backslash\calVW_{l-1}$, i.e. $0<z_l<|V_l|-1$.
One easily finds that $\Np{v^\prime}{v_{l}^0}=1$ and $|(v^\prime\to v_l^0)|=2$, which implies that $\{\csd{v_{l}^0},\as{v^\prime}\}=-1$. 
Examining the assumption (A1), we can also find a vertex $v\in\calVW_{l-1}$ 
whose corresponding $\as{v}$ satisfies
\begin{equation}
 \{\csd{v_{l}^0},\as{v}\}=\sum_{j=1}^{\Np{v}{v_l^0}}(-1)^{|(v\to v_l^0)|-1} \ne 0. 
\end{equation}    
Here we used the assumption that $\Np{v}{v_l^0}$ is odd. 
This implies that $\as{v}$ and $\as{v^\prime}$ are directly connected at the vertex $v_l^0$, 
and thus $a$ operators $\{\as{v}\}_{v\in\calVW_{l}}$ are connected if $\{\as{v}\}_{v\in\calVW_{l-1}}$ are connected.

Since the $a$ operators $\{\as{v}\}_{v\in\calVW_1}$ are apparently connected at the vertex $v_1^0$, 
we conclude that the $a$ operators $\{\as{v}\}_{v\in\calVW_L}$ are connected.
This completes the proof of Proposition~\ref{proposition3}.   
\section{Further extensions}
\label{s:Further extensions}
Here we consider further extensions.
Let $\sfS=[\sfS_{l,l^\prime}]_{l,l^\prime=1}^L$ be a positive definite $L\times L$ matrix. 
Then, by using the $b$ operators defined in \eqref{eq:b-operator} and $\sfS$, we define the hopping Hamiltonian
\begin{equation}
 \Hhop(\sfS) = \sum_{l,l^\prime=1}^L \sum_{\sigma=\up,\dn} \sfS_{l,l^\prime} \bsd{l}\bs{l^\prime} 
\end{equation}
on the graph $\calG_L$.
Note that $\Hhop(\sfS)$ with $\sfS_{l,l^\prime}=t\delta_{l,l^\prime}$ is reduced to $\Hhop$. 
Since $\Hhop(\sfS)$ is a positive semidefinite operator and is made up of only the $b$ operators, 
the $a$ operators defined in \eqref{eq:a-operator} still correspond to the single-electron zero-energy states of $\Hhop(\sfS)$.%
\footnote{
Since $\sfS$ is positive, the dimension of the single-electron zero-energy states remains $|\calV|-L$.
}
Therefore Propositions~\ref{proposition1}, \ref{proposition2} and \ref{proposition3} hold true even if we
replace $\Hhop$ with $\Hhop(\sfS)$.

Next, we consider a deformation of the $b$ operators. Let us define 
\begin{equation}
 \Bs{l} = \sum_{v\in V_l} \lambda_l(v) \cs{v}
\end{equation}
where $\lambda_l(v)$ are complex numbers and $\lambda_l(v_l^0)=1$. The operator $\Bs{l}$ corresponds to 
a localized state on the subset $V_l$ of $\calV_L$.  
By using the $B$ operators we define the hopping Hamiltonian
\begin{equation}
  \Hhop^\prime(\sfS) =\sum_{l,l^\prime=1}^L \sum_{\sigma=\up,\dn}\sfS_{l,l^\prime}\Bsd{l}\Bs{l^\prime} 
 \label{eq:Hhopprime}
\end{equation}
with positive definite matrix $\sfS$.
As in the case of the $b$ operators, we can show the linear independence of the $B$ operators.
Therefore, Propositions~\ref{proposition1} and ~\ref{proposition2} with $\Hhop$ 
replaced by $\Hhop^\prime(\sfS)$ hold true.  

Let us consider a set of fermion operators which anticommute with the $B$ operators. 
Let $v$ be a vertex in $\calVW_L$ and let $(v\to u)=(v_0,\dots,v_k)$ be a directed path in $\calGd_L$. 
Recall that each vertex except $v=v_0$ in $(v_0,v_1,\dots,v_{k})$ is one of the black vertices in the complete graphs 
$\{G_l\}_{l=1}^L$. 
Recall also that each $v_{j-1}$ with $j=2,\dots,k$ is merged with a white vertex of $G_{l(v_j)}$, where $l(v_j)=l$ if $v_j=v_l^0$.
We then define 
\begin{equation}
 \lambda[(v\to u)]
  =
  \lambda[(v_0,\dots,v_k))]=
  \left\{
   \begin{array}{@{\,}ll}
    1 & \mbox{if $u=v$}\\
    \displaystyle
  \prod_{j=1}^k (-\lambda_{l(v_{j})}(v_{j-1}))
      & \mbox{otherwise},
       \end{array}
	\right.
\end{equation}
and 
\begin{equation}
 \Asd{v}= \sum_{u\in R(v)}
  \left(
   \sum_{j=1}^{\Np{v}{u}}
   \lambda[(v\to u)_j)]
  \right)
  \csd{u}.
\end{equation}
We can verify that the $A$ operators anticommute with the $B$ operators. In fact, we have
\begin{eqnarray}
 \{\Asd{v},\Bs{l}\}
&=&\sum_{u\in R(v)\cap V_l}
 \left(\sum_{j=1}^{\Np{v}{u}}\lambda[(v\to u)_j]\lambda_l(u)\right)
\nonumber\\
&=&
 \sum_{u\in (R(v)\cap V_l)\backslash\{v_l^0\}}
 \left(\sum_{j=1}^{\Np{v}{u}}\lambda[(v\to u)_j]\lambda_l(u)\right)
 + 
 \left(\sum_{j=1}^{\Np{v}{v_l^0}}\lambda[(v\to v_l^0)_j]\lambda_l(v_l^0)\right)
\nonumber\\ 
&=&
 \sum_{u\in (R(v)\cap V_l)\backslash\{v_l^0\}} 
  \left(
   \sum_{j=1}^{\Np{v}{u}}
   \left(
    \lambda[(v\to u)_j]\lambda_l(u)
    +
    \lambda[(v\to u)_j](-\lambda_{l(v_l^0)}(u))
   \right)
  \right)
\nonumber\\
&=& 0.
\end{eqnarray}
Here we used $\lambda_l(v_l^0)=1$ and  $\lambda[(v\to v_l^0)_j] = \lambda[(v\to u)_j](-\lambda_{l(v_l^0)}(u))$
for $(v\to v_l^0)=(v,\dots,u,v_l^0)$.

The uniqueness of the ferromagnetic ground states of $\Hhop^\prime(\sfS)+\Hint$ with $\Ne=|\calV_L|-L$ follows from 
the connectivity of $\{\As{v}\}_{v\in\calVW_L}$.
It is easy to see that the following assumption is sufficient for the connectivity of $\{\As{v}\}_{v\in\calVW_L}$.
\bigskip\\\noindent
(A2) 
For every $l\in \{2,\dots,L\}$ such that $z_l\ne|V_l|-1$, 
there exists a white vertex $v\in\calV_{l-1}^\rmW$ 
for which 
${\sum_{j=1}^{\Np{v}{v_l^0}}\lambda[(v\to v_l^0)_j)]}$
is non-vanishing.
\bigskip\\\noindent
Proposition~\ref{proposition3} with (A1) and $\Hhop$ replaced by (A2) and $\Hhop^\prime(\sfS)$ 
remains to hold.  
\jour{ 
\bigskip\\
{\small
\textbf{Acknowledgements}~~
I would like to thank H. Katsura and H. Tasaki for valuable discussions.
I also would like to thank P. M\"uhlbacher, V. Sohinger and  D. Ueltschi for their warm hospitality and fruitful discussions 
during my stay at the Mathematics Institute of the University of Warwick,
where a part of this work was performed.}
}
{ 
\begin{acknowledgements}
I would like to thank H. Katsura and H. Tasaki for valuable discussions.
I also would like to thank P. M\"uhlbacher, V. Sohinger and  D. Ueltschi for their warm hospitality and fruitful discussions 
during my stay at the Mathematics Institute of the University of Warwick,
where a part of this work was performed.
\end{acknowledgements}
}
\begin{appendices}
\renewcommand{\theequation}{\thesection.\arabic{equation}}
\renewcommand{\thefigure}{\thesection.\arabic{figure}}
\setcounter{equation}{0}
\setcounter{figure}{0}
\section{Flat-band models on line graphs of \\2-connected graphs}
\label{s:appendixA}
\subsection{Line graphs and cell construction}
\label{s:LineGraph}
The problem of flat-band ferromagnetism on line graphs was studied and fully resolved by 
Mielke~\cite{Mielke91a,Mielke91b,Mielke92}.
In this appendix we revisit the problem using our method.

\begin{figure}
 \begin{center}
     \includegraphics[width=.9\textwidth]{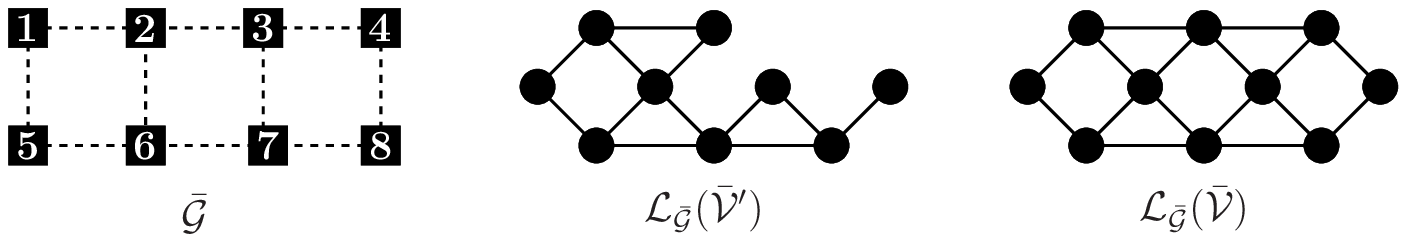}
 \end{center}
\caption{
We associate 2-vertex complete graphs and 3-vertex complete graphs, respectively, with vertices 1,4,5,8 and 2,3,6,7
in $\bcalG$. The figure in the middle is $\LG(\bcalV^\prime)$ with $\bcalV^\prime = \{1,2,5,6,7,8\}$. 
There are three boundary vertices in $\LG(\bcalV^\prime)$ 
and subgraphs $G_{2}, G_7$ and $G_8$ are in the boundary of $\LG(\bcalV^\prime)$.
The graph $\LG(\bcalV)$ is the line graph of $\bcalG$.
}
\label{fig:example-bG}
\end{figure}

Let $\bcalG=(\bcalV,\bcalE)$ be a simple 2-connected graph.%
\footnote{
An edge connecting a vertex to itself is called a loop.
A simple graph has no loops or multiple edges.
A vertex is called a cut vertex if its removal disconnects the connected graph.  
A 2-connected graph has no cut vertices. 
}
With each $\bv\in\bcalV$ we associate a complete graph $G_\bv=(V_\bv,E_\bv)$, 
where $|V_\bv|$ is equal to the number of edges touching $\bv$, and
\begin{equation}
  E_\bv=\{e=\{v,v^\prime \}~|~v,v^\prime \in V_\bv, v\ne v^\prime \}.  
\end{equation}
Let $\bcalV^\prime$ be a subset of $\bcalV$. From the collection $\{G_\bv\}_{\bv\in\bcalV^\prime}$ 
of complete graphs we create a new graph $\LG(\bcalV^\prime)$ in the following way.
For all $\bv\in\bcalV^\prime$,
we arrange $G_\bv$ at $\bv$ in such a way that every vertex of $G_\bv$ 
is placed at each one of the edges touching $\bv$. 
If we find two vertices on the same edge, we identify them as a single vertex. 
Then we regard the resulting set of vertices as the vertex set of $\LG(\bcalV^\prime)$.
The edge set of $\LG(\bcalV^\prime)$
is given by $\sum_{\bv\in\bcalV^\prime}E_\bv$ with the above vertex identification.
We say that a vertex in the graph $\LG(\bcalV^\prime)$ 
has weight 2 if it comes from two complete graphs and weight 1 otherwise. 
We use $G_\bv$ to denote also the subgraph of $\LG(\bcalV^\prime)$
whose vertices and edges come from the complete graph $G_\bv$.
A vertex with weight 1 is called  a boundary vertex of $\LG(\bcalV^\prime)$. 
If a subgraph $G_\bv$ contains a boundary vertex, 
we say that $G_\bv$ is in the boundary of $\LG(\bcalV^\prime)$. 
In the case $\bcalV^\prime = \bcalV$,
the graph $\LG(\bcalV)$ obtained by the above procedure is nothing but the line graph of $\bcalG$,
since we find a vertex in $\LG(\bcalV)$ on every edge in $\bcalE$ and there exists in $\LG(\bcalV)$ 
an edge $\{v,v^\prime\}$ which connects $v$ on $e\in\bcalE$ and $v^\prime$ on $e^\prime\in\bcalE$
if and only if $e$ and $e^\prime$ touch the same vertex in $\bcalV$.
We remark that, since $\bcalG$ is 2-connected, the weight of every vertex in $\LG(\bcalV)$ is 2, i.e., 
$\LG(\bcalV)$ has no boundary vertices.
See Fig.~\ref{fig:example-bG} for an example. 

The above way to create a line graph as an assembly of complete graphs 
indicates that we may treat the flat-band models on line graphs 
by using our method with $\{G_\bv\}_{\bv\in\bcalV}$.
It is, however, impossible to use our construction procedure  
to make the coloured graph corresponding to the line graph $\LG(\bcalV)$ directly, 
since the black vertex in the newly added complete graph can not be identified 
with a vertex in the so far constructed coloured graph, 
which means that the weight of the black vertex $v_L^0$ is one. 
(Recall that $\LG(\bcalV)$ has no boundary vertices.)
In fact, we can instead construct the coloured graph whose structure is 
almost the same as $\LG(\bcalV)$ through our method.
We first consider the flat-band model on this almost the same graph, 
and then return to the problem of flat-band ferromagnetism on the correct line graph.
         
\begin{figure}
 \begin{center}
     \includegraphics[width=.9\textwidth]{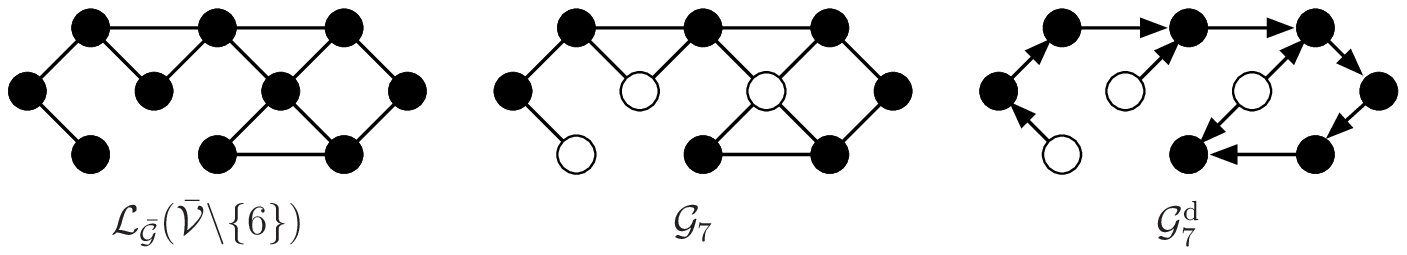}
 \end{center}
\caption{
Setting $(\bv_1,\bv_2,\dots,\bv_7)=(5,1,2,3,4,8,7)$ for $\bcalG$ in Fig.\ref{fig:example-bG}, we can construct 
the coloured graph $\calG_7=(\LG(\bcalV\backslash\{6\}),\gamma_7)$ 
as depicted in the middle. 
}
\label{fig:example-bG2}
\end{figure}

Choose one vertex, which we denote by $\bu$, in $\bcalV$ and consider $\LG(\bcalV\backslash\{\bu\})$. 
The coloured graph corresponding to $\LG(\bcalV\backslash\{\bu\})$ 
can be constructed from $\{G_\bv\}_{\bv\in\bcalV\backslash\{\bu\}}$ 
through our procedure described in section~\ref{s:Definition}.
This claim is shown as follows.
Let $\bv_1,\bv_2,\dots,\bv_L$  be a sequence of vertices in $\bcalV\backslash\{\bu\}$,
where $L=|\bcalV|-1$ and $\bv_l\ne\bv_{l^\prime}$ for $l\ne l^\prime$.
We denote by $\bcalV_l$ the set of vertices $\{\bv_1,\bv_2,\dots,\bv_l\}$.
As we will see below, it is possible to choose a sequence $\bv_1,\bv_2,\dots,\bv_L$ so that
a sequence of graphs 
$\LG(\bcalV_1),\LG(\bcalV_2),\dots,\LG(\bcalV_{L})$
may satisfy the following properties;
$\LG(\bcalV_l)$ is connected for all $l=1,\dots,L$ 
and a sequence $\calV_1,\dots,\calV_L$, where $\calV_l$ denotes the edge set of $\LG(\bcalV_l)$, is strictly increasing, i.e. 
$\calV_{l}\backslash\calV_{l-1}\ne\emptyset$ for all $l=2,\dots,L$.
Once we find a sequence $\bv_1,\bv_2,\dots,\bv_L$ for which the above properties hold, 
it is possible to construct coloured graphs $\calG_1,\calG_2,\dots,\calG_L$  
such that the vertex and the edge sets of $\calG_l$ are given by those of $\LG(\bcalV_l)$ in the following manner.
For $\bv\in\bcalV\backslash\{u\}$, paint one of the vertices in $G_{\bv}$ black 
and paint all the other vertices white. 
We first arrange $G_{\bv_1}$ at $\bv_1\in\bcalV$ placing every vertex at each one of the edges touching $\bv_1$ 
and set $\calG_1=(G_{\bv_1},\gamma_1)=(V_{\bv_1},E_{\bv_1},\gamma_1)$.
Then, for $l=2,\dots,L$, we arrange $G_{\bv_l}$ at $\bv_l\in\bcalV$ 
placing every vertex at each one of the edges touching $\bv_l$ 
and identify two vertices on the same edge
to merge $G_{\bv_l}$ with the so far constructed coloured graph $\calG_{l-1}$.
We note that there always exist, in $\calV_{l-1}$, vertices which can be identified 
with white vertices in $G_{\bv_l}$ since $\LG(\bcalV_{l})$ is connected. 
We also note that it is always possible to place the black vertex in $G_{\bv_l}$ on an edge in $\bcalE$ 
since $\calV_l\backslash\calV_{l-1}$ is non-empty.  
This completes the proof of the claim. 
(See Fig.~\ref{fig:example-bG2} for an example corresponding to $\bcalG$ in Fig.~\ref{fig:example-bG}.)

Let us check that we can always find a sequence $\bv_1,\dots,\bv_L$ for which the above properties are satisfied. 
We first note that $\LG(\bcalV_{L})=\LG(\bcalV\backslash\{\bu\})$ is connected since  $\bcalG$ 
is 2-connected and
that there are $|V_{\bu}|$ boundary vertices in $\LG(\bcalV_{L})$.
Now suppose that $\LG(\bcalV_l)$ is connected and has at least one boundary vertex. 
Among subgraphs in the boundary of $\LG(\bcalV_{l})$  
we can select one subgraph, which we label as $G_{\bv_{l}}$, 
so that the graph $\LG(\bcalV_{l}\backslash\{\bv_l\})$ is connected.
To see this, assume that any graph $\LG(\bcalV_{l}\backslash\{\bv\})$
where $G_{\bv}$ is in the boundary of $\LG(\bcalV_l)$ is disconnected.
This assumption implies that there exists a vertex $\bv$ 
for which $\LG(\bcalV_{l}\backslash\{\bv\})$ can be decomposed as
$\LG(\bcalV_{l}\backslash\{\bv\})=\LG(\bcalV_{l}^1)\cup\LG(\bcalV_{l}^2)$, 
where 
$\LG(\bcalV_{l}^1)$ is disconnected with $\LG(\bcalV_{l}^2)$, 
in such a way that at least one of $\LG(\bcalV_{l}^1)$ and  $\LG(\bcalV_{l}^2)$, say $\LG(\bcalV_{l}^1)$, is connected and  
has no boundary vertices except ones which are originally contained in the subgraph $G_\bv$ of $\LG(\bcalV_{l})$.%
\footnote{
This will be proved as follows.
Let $\calV_l^\mathrm{b}$ be the set of boundary vertices of $\LG(\bcalV_{l})$.
We pick up a vertex in $\calV_l^\mathrm{b}$ and 
remove the subgraph $G_{\bv^\prime}$ containing this boundary vertex from $\LG(\bcalV_{l})$.
Then the resulting graph $\LG(\bcalV_{l}\backslash\{\bv^\prime\})$ is decomposed into connected graphs.
Among the connected graphs we select one which has the least number of boundary vertices in $\calV_l^\mathrm{b}$.  
Let $\calV_{l,1}^\mathrm{b}$ be a subset of $\calV_{l}^\mathrm{b}$ whose elements are in the selected graph.
We pick up a vertex in $\calV_{l,1}^\mathrm{b}$ and remove the subgraph $G_{\bv^{\prime\prime}}$
containing this boundary vertex from $\LG(\bcalV_{l})$. 
The resulting graph $\LG(\bcalV_{l}\backslash\{\bv^{\prime\prime}\})$ is again decomposed into connected graphs. 
Let $\calV_{l,2}^\mathrm{b}$ be a subset of $\calV_{l}^\mathrm{b}$ whose elements are in the connected graph
having the least number of boundary vertices in $\calV_l^\mathrm{b}$.    
Here note that $\calV_{l,2}^\mathrm{b}\subset \calV_{l,1}^\mathrm{b} \subset \calV_{l}^\mathrm{b}$ and
the number of elements in $\calV_{l,2}^\mathrm{b}$ is strictly less than that in $\calV_{l,1}^\mathrm{b}$.
Repeating the same procedure we reach the claim. 
}
Considering the implications of this fact for the graph $\bcalG$,
we find that there are no edges in $\bcalE$ which connect vertices in $\bcalV_l^1$ 
and vertices except $\bv$ in 
$\bcalV\backslash\bcalV_l^1$, i.e., $\bv$ is a cut vertex in $\bcalG$.%
\footnote{
Assume that there is an edge $\{\bv^\prime, \bv^{\prime\prime}\}$ with 
$\bv^\prime\in \bcalV_l^1$ and $\bv^{\prime\prime}\in\bcalV\backslash(\bcalV_l^1\cup\{\bv\})$. 
Since the vertex at the edge $\{\bv^\prime, \bv^{\prime\prime}\}$ is not a boundary vertex of $\LG(\bcalV_{l}^1)$, 
it has weight 2. 
This implies that the subgraphs $G_{\bv^\prime}$ and $G_{\bv^{\prime\prime}}$ are in $\LG(\bcalV_{l}\backslash\{\bv\})$,
which contradicts that $\LG(\bcalV_{l}^1)$ is disconnected with $\LG(\bcalV_{l}^2)$. 
} 
This, however, contradicts with the assumption that $\bcalG$ is 2-connected.         
Therefore, we can find a vertex $\bv_l\in\bcalV_l$ for which $\LG(\bcalV_{l}\backslash\{\bv_l\})$ is connected.
We set $\bcalV_{l-1}=\bcalV_{l}\backslash\{\bv_l\}$. It is obvious that $\calV_l\backslash\calV_{l-1}$ is non-empty since
the subgraph $G_{\bv_l}$ of $\LG(\bcalV_{l})$ is in the boundary and contains boundary vertices in $\calV_l\backslash\calV_{l-1}$. 
It is also obvious that $\LG(\bcalV_{l-1})$ has at least one
boundary vertex, since there exists a vertex with weight 2 in the subgraph $G_{\bv_l}$ of 
$\LG(\bcalV_{l})$ and
this vertex becomes to have weight 1 in $\LG(\bcalV_{l-1})$.
Therefore, we can inductively find vertices $\bv_L,\bv_{L-1},\dots,\bv_1$ 
for which $\{\LG(\bcalV_{l})\}_{l=1}^L$ satisfies the desired properties.   
\subsection{Flat-band models on $\LG(\bcalV\backslash\{\bu\})$}
\label{s:almost line graph}
Let us consider the Hubbard Hamiltonian $H=\Hhop+\Hint$ on the coloured graph 
$\calG_L=(\LG(\bcalV\backslash\{\bu\}),\gamma_L)=(\calV_L,\calE_L,\gamma_L)$
where
\begin{equation}
 \Hhop=t \sum_{\bv\in\bcalV\backslash\{\bu\}} \sum_{\sigma=\up,\dn} \bsd{\bv} \bs{\bv}
 \label{eq:Hhop on line graphs}
\end{equation}
 and
\begin{equation}
 \bs{\bv} = \sum_{v\in V_\bv} \cs{v}.
\end{equation}
From proposition~\ref{proposition1} we find that the degeneracy of 
the single-electron ground states of $\Hhop$ is 
$|\calV_L|-L=|\bcalE|-|\bcalV|+1$.%
\footnote{
We have $|\calV_L|=|\bcalE|$ since the number of vertices in $\calV_L$ is equal to 
that of edges in $\bcalE$.  
}
Note that the degeneracy is independent of whether the graph $\bcalG$ is bipartite or not in this case.%
\footnote{
As is well known, for the Hubbard Hamiltonian 
on the line graph $\LG(\bcalV)$ of $\bcalG$,
the degeneracy is $|\bcalE|-|\bcalV|+1$ if $\bcalG$ is bipartite and $|\bcalE|-|\bcalV|$ otherwise.
}

Let us show that the assumption (A1) is satisfied for the directed coloured graph $\calGd_L$ corresponding to $\calG_L$ 
and thus the ground state of $H$ on the graph $\LG(\bcalV\backslash\{\bu\})$ 
with $\Ne=|\bcalE|-|\bcalV|+1$ has the unique ferromagnetic 
ground states.  

Before proceeding to the proof, we comment on the properties of $\calGd_L$ which are characteristic to the present case.
Firstly, since the weight of each vertex in $\calGd_L$ is at most 2, each white vertex has at most 2 directed paths which end 
at black vertices with weight 1. 
Secondly, two subsets $V_\bv$ and $V_{\bv^\prime}$ share at most one vertex since the graph $\bcalG$ is simple.
The second fact implies that there is at most one directed edge which starts from a black vertex. 

We shall show by induction that, 
for all $l=2,\dots,L$, there exists a white vertex $v\in\calVW_{l-1}$ from which $v_l^0\in\calVB_L$ is reachable 
by a single directed path, i.e., for which we have $\Np{v}{v_l^0}=1$.  
To begin with, consider the black vertex $v_l^0$ with $l=2$. Since $V_{\bv_1}$ and $V_{\bv_2}$ share exactly one vertex, 
we can clearly find, in $\calVW_1$, a white vertex reachable to $v_2^0$ by a single directed path. 
Let $m$ be an integer in $3\le m \le L$ and assume that the above claim is true for all $l=2,\dots,{m-1}$.  
Under this assumption,  
assume also that there are no white vertices from which $v_m^0$ is reachable by a single directed path.
As we will show in the following, the second assumption leads to a contradiction,
and therefore we obtain the desired claim.
          
Consider the construction step of $\calG_m$ and $\calGd_m$.
If white vertices of $\calG_{m-1}$ only are identified with white vertices of $G_{\bv_m}$, 
one can pick up, among those identified vertices, a white vertex $v\in\calVW_{m-1}$ from which $v_m^0$ 
is reachable by a single directed path $(v\to v_m^0)=(v,v_m^0)$.
Since this is against the second assumption,  
at least one black vertex $v_{m_1}^0$ with $m_1\le {m-1}$ must be identified with a white vertex of $G_{\bv_m}$.  
By the inductive assumption we then find, in $\calVW_{m_1-1}$, a white vertex $v^\prime$ which is reachable  
to $v_{m_1}^0$ by a single directed path.
Now note that $v^\prime$ is also reachable to $v_{m}^0$ by the directed path which first reaches 
$v_{m_1}^0$ then goes to $v_{m}^0$.
Therefore, by the second assumption, there must be another directed path connecting $v^\prime$ with $v_{m}^0$.
Since a black vertex $v_{m_2}^0\in\calV_{m_1-1}$ which is connected with $v^\prime$ 
by a directed edge $(v^\prime,v_{m_2}^0)$ must be included in either of the two directed paths starting from $v^\prime$,
the above observation implies that we have the black vertex $v_{m_2}^0$ with $m_2<m_1$ which is reachable to $v_m^0$.
By the assumptions, we again find, 
in $\calVW_{m_2-1}$, a white vertex $v^{\prime\prime}$ which is reachable not only to $v_{m_2}^0$ by a single directed path 
but also to $v_{m}^0$ by two directed paths. Repeating the same argument as above we finally conclude that $v_1^0$
is reachable to $v_m^0$ by a single path and
every white vertex in 
$\calVW_1=V_{\bv_1}$ is reachable to $v_m^0$ by two directed paths.  
As we will see below, this furthermore implies that every black vertex $v_l^0$ with $l=1,\dots,{m-1}$ is reachable to
$v_m^0$ by a single directed path and every white vertex in $\calVW_{m-1}$ is reachable to $v_m^0$ by two directed paths.   
As a consequence of this, we find that any vertex in $\calV_{m-1}$ is shared by exactly two subsets 
in $V_{\bv_1},\dots,V_{\bv_m}$, i.e., the weight of any vertex in $\calV_{m-1}$ is 2.
The implication of this for the graph $\bcalG$ is
that the vertex $\bv_m\in\bcalV$ at which we arrange $G_{\bv_m}$ is a cut vertex, 
which contradicts that $\bcalG$ is 2-connected.

Under the assumptions, let us show that, 
if every white vertex in $\calVW_1$ is reachable to $v_m^0$ by two directed paths,
every black vertex $v_l^0$ with $l=1,\dots,{m-1}$ is reachable to
$v_m^0$ by a single directed path and every white vertex in $\calVW_{m-1}$ is reachable to $v_m^0$ by two directed paths.
Let $u_{1}^{1},u_{1}^{2}\dots,u_{1}^{r_1}$ with $r_1=|\calVW_1|$ be 
white vertices in $\calVW_{1}$ and let $(u_{1}^{k}\to v_m^0)_j$ with $j=1,2$ 
be directed paths which start from $u_{1}^{k}$.
We note that all the directed paths starting from $u_1^k$ end at $v_m^0$,
since there is at most one directed edge which starts from a black vertex.
If there are no white vertices reachable to any black vertex in $(u_{1}^k\to v_m^0)_j$ with $k=1,\dots,r_1$ and $j=1,2$ 
in $\calV_{m-1}$, the above claim is apparently true.
Now supposing that there are, we denote by $u_{2}^{1},u_{2}^{2},\dots,u_{2}^{r_2}$ the white vertices in $\calVW_{m-1}$ 
which are reachable to black vertices in $(u_1^k\to v_m^0)_j$ with $k=1,\dots,r_1$ and $j=1,2$.   
Recall that we are assuming 
that there are no white vertices reachable to $v_m^0$ by a single directed path.
Since $u_{2}^{1},u_{2}^{2},\dots,u_{2}^{r_2}$ are reachable to $v_m^0$ 
by directed paths each of which joins one of $(u_1^k\to v_m^0)_j$,
there must be another directed path starting from each  $u_{2}^{k}$,
which we denote by $(u_{2}^{k}\to v_m^0)_2$.
If there are no white vertices reachable to any black vertex in $(u_{2}^{k}\to v_m^0)_2$ with $k=1,\dots,r_2$,
we conclude that the above claim is true.
If there are, repeating the above argument, we eventually reach the conclusion that the claim is true.
\subsection{Flat-band models on line graphs of bipartite graphs}
In this section, we consider the Hubbard Hamiltonian on the line graph $\LG(\bcalV)$ 
of a bipartite graph $\bcalG=(\bcalV,\bcalE)$. 
Let us assume that $\bcalV$ is decomposed as $\bcalV=\bcalA\cup\bcalB$  where $\bcalA\cap\bcalB=\emptyset$.
Without loss of generality we assume that $\bu$ is in $\bcalA$. It is easy to see 
that the $b$ operator 
\begin{equation}
 \bs{\bu}=\sum_{v\in V_{\bu}}\cs{v}
\end{equation}
is represented as
\begin{equation}
 \bs{\bu}=\sum_{v\in\bcalB}\bs{v}-\sum_{v\in\bcalA\backslash\{\bu\}}\bs{v}.
\end{equation}
The Hubbard Hamiltonian on the line graph $\LG(\bcalV)$ is written as
$H=\Hhop(\sfS)+\Hint$
with
\begin{equation}
\Hhop(\sfS) = t \sum_{\bv\in\bcalV} \sum_{\sigma=\up,\dn} \bsd{\bv}\bs{\bv}
      =\sum_{\bv,\bv^\prime\in\bcalV\backslash\{\bu\}} 
       \sum_{\sigma=\up,\dn} \sfS_{\bv,\bv^\prime} \bsd{\bv}\bs{\bv^\prime}
\end{equation}
where $\sfS_{\bv,\bv^\prime}=2t$ if $\bv=\bv^\prime$, $\sfS_{\bv,\bv^\prime}=t$ if $\bv$ and $\bv^\prime$ are in the same sublattice, and
$\sfS_{\bv,\bv^\prime}=-t$ if $\bv\in\bcalA,\bv^\prime\in\bcalB$ or $\bv\in\bcalB,\bv^\prime\in\bcalA$.%
\footnote{
This Hamiltonian and \eqref{eq:HamiltonianMeilke} have a uniform on-site potential 
and are equivalent to that treated by Mielke in~\cite{Mielke91a,Mielke91b,Mielke92}.
}
Then, from the results in section~\ref{s:Further extensions} and appendix~\ref{s:almost line graph}, 
we conclude that the Hubbard Hamiltonian $H=\Hhop(\sfS)+\Hint$ on the line graph 
of a bipartite graph $\bcalG=(\bcalV,\bcalE)$  
has the unique ferromagnetic ground states if $\Ne=|\bcalE|-|\bcalV|+1$.
\subsection{Flat-band models on line graphs of non-bipartite graphs}
In this section, we consider the Hubbard Hamiltonian on the line graph $\LG(\bcalV)$ 
of a non-bipartite graph $\bcalG=(\bcalV,\bcalE)$. 
Let us consider the Hubbard Hamiltonian 
\begin{equation}
 H=\Hhop + t \bsd{\bu}\bs{\bu} + \Hint
\label{eq:HamiltonianMeilke}
\end{equation}
with $\Hhop$ in \eqref{eq:Hhop on line graphs}.
In this case single-electron ground states for $H$ must be the zero-energy states for $t \bsd{\bu}\bs{\bu}$
in addition to $\Hhop$. 
Note that we have already known from the result in appendix~\ref{s:almost line graph} that $a$ operators given in \eqref{eq:a-operator} correspond to
the single-electron zero-energy states for $\Hhop$.
Therefore, our task is to form linear combinations of $a$ operators so that they may anticommute with $\bsd{\bu}$.
Here we note that $b$ operators $\{\bs{\bv}\}_{\bv\in\bcalV}$ are linearly independent unlike the bipartite case 
and thus the dimension of the single-electron zero-energy states is $|\calV_L|-(L+1)=|\bcalE|-|\bcalV|$.
This means that there exists at least one $a$ operator which does not anticommute with $\bsd{\bu}$.%
\footnote{
If all the $a$ operators anticommuted with $\bsd{\bu}$, we would have $|\bcalE|-|\bcalV|+1$ single-electron zero-energy states.
}
We pick up one and denote by $\as{\tilde{v}}$ this $a$ operator.
Then, for each $v\in\calVW_L\backslash\{\tilde{v}\}$, we define
\begin{equation}
\label{eq:tilde a}
 \tas{v}=\as{v}-\frac{\left\{ \asd{v}, \bs{\bu} \right\}} 
                      {\left\{ \asd{\tilde{v}}, \bs{\bu} \right\}}\as{\tilde{v}} . 
\end{equation} 
It is easy to check that $\{\tasd{v}, \bs{\bv}\}=0$ for any 
$v\in\calVW_L\backslash\{\tilde{v}\}$ and any $\bv\in\bcalV$.
The $\tilde{a}$ operators are linearly independent since the coefficients of $c_{v,\sigma}$ 
in the expression~\eqref{eq:tilde a} is non-vanishing only for $\tas{v}$. 
Therefore, if $\{\tas{v}\}_{v\in\calVW_L\backslash\{\tilde{v}\}}$ is connected, 
$H=\Hhop + t \bsd{\bu}\bs{\bu} + \Hint$ with $\Ne=|\bcalE|-|\bcalV|$ has
the unique ferromagnetic ground states, which are give by 
$\prod_{v\in\calVW_L\backslash\{\tilde{v}\}} \left(\tilde{a}_{v,\up}^\dagger\right)\Phi_0$
and its SU(2) rotations.
Unfortunately, it is not easy to obtain a sufficient condition for $\bcalG$ under which 
$\{\tas{v}\}_{v\in\calVW_L\backslash\{\tilde{v}\}}$ is connected.
In fact, as was remarked by Mielke in Ref.~\cite{Mielke92}, 2-connectedness of $\bcalG$ 
is not sufficient to prove the connectivity of $\{\tas{v}\}_{v\in\calVW_L\backslash\{\tilde{v}\}}$. 
It seems to be easier to check the connectivity for each example of concrete models, 
so that we do not pursue this problem further here.  
\section{Proof of the uniqueness of the ferromagnetic ground states}
\label{s:appendixB}
\setcounter{equation}{0}
In this appendix, just for the sake of the readers' convenience,
we shall prove the uniqueness of the ferromagnetic ground states 
under the assumption of the connectivity of $\{\as{v}\}_{v\in\calVW_L}$
within the present notations.  

We suppose that $\Ne=|\calV_L|-L=|\calVW_L|$ 
and $\{\as{v}\}_{v\in\calVW_L}$ is connected.
Since $\Hhop$ and $\Hint$ are positive semidefinite, the eigenvalues of $H$ are greater than or equal to zero.
When $\Ne=|\calVW_L|$, we know that $\Phi_\up$ in \eqref{eq:ferro} is a zero energy state for
both $\Hhop$ and $\Hint$.
Thus the ground state energy of $H$ is zero, 
and any ground state of $H$ must be a zero-energy state of both $\Hhop$ and $\Hint$.

Let $\Phi$ be a ground state of $H$. 
Considering the expansion of $\Phi$ in terms of $a$ and $b$ operators,
one finds that $\Phi$, which is a zero-energy state of $\Hhop$, is expressed as
\begin{equation}
 \Phi=\sum_{V_\up,V_\dn\subset\calVW_L;|V_\up|+|V_\dn|=\Ne} \phi(V_\up;V_\dn) 
\left(\prod_{v\in V_\up}a_{v,\up}^\dagger\right)
\left(\prod_{v\in V_\up}a_{v,\dn}^\dagger\right)\Phi_0
\end{equation} 
where $\phi(V_\up;V_\dn)$ are real coefficients. 
A ground state $\Phi$ in this expression must also satisfy $\Hint\Phi=0$.
This condition is reduced to $c_{v,\dn} c_{v,\up}\Phi=0$ for all $v\in\calV_L$
since $\Hint$ is a sum of the positive semidefinite operators 
$Un_{v,\up} n_{v,\dn}=Uc_{v,\up}^\dagger c_{v,\dn}^\dagger c_{v,\dn} c_{v,\up}$.
Firstly examining $c_{v,\dn} c_{v,\up}\Phi=0$ for all $v\in\calVW_L$, 
we find that $\phi(V_\up;V_\dn)=0$ if $V_\up\cap V_\dn\ne\emptyset$. 
We thus obtain the expression of $\Phi$
\begin{equation}
 \Phi=\sum_{\bm{\sigma}}\phi(\bm{\sigma}) 
  \left(\prod_{v\in \calVW_L} a_{v,\sigma_{v}}^\dagger\right)\Phi_0,
\end{equation}
where $\phi(\bm{\sigma})$ are real coefficients,
$\bm{\sigma}=(\sigma_v)_{v\in\calVW_L}$ with $\sigma_v=\up,\dn$ 
is a spin configuration of $a$ operators
and the summation is taken over all spin configurations.

Let us examine the remaining conditions $c_{v,\dn} c_{v,\up}\Phi=0$ for all $v\in\calVB_L$.
Now suppose that $a_{v,\sigma}$ and $a_{v^\prime}$ are directly connected at a black vertex $v_l^0$.%
\footnote{
Recall that $a$ operators are connected at black vertices.
}
Then, for any spin configuration $\bm{\tau}$,
we obtain from $c_{v_l^0,\dn} c_{v_l^0,\up}\Phi=0$ that 
\begin{equation}
\left(\prod_{u\in\calVW_L;u\ne v,v^\prime} c_{u,\tau_u} \right)c_{v_l^0,\dn} c_{v_l^0,\up}\Phi
 ={\bf sgn}[v,v^\prime]
 \alpha_{v}(v_l^0)\alpha_{v^\prime}(v_l^0)
 \left( \phi(\bm{\tau}_{{\tau_v=\up,\tau_{v^\prime}=\dn}})-\phi(\bm{\tau}_{\tau_v=\dn,\tau_{v^\prime}=\up}) \right)
 =0
\end{equation}
where 
$\alpha_v(u)$ is the coefficient of $c_{u,\sigma}$ in \eqref{eq:a-operator},
${\bf sgn}[v,v^\prime]$ is a sign factor arising from the exchange of $a$ operators, and
$\bm{\tau}_{\tau_v=\sigma,\tau_{v^\prime}=\sigma^\prime}$ is a spin configuration which is obtained from
$\bm{\tau}$ by replacing $\tau_v$ and $\tau_{v^\prime}$ with $\sigma$ and $\sigma^\prime$, respectively. 
We thus have 
$\phi(\bm{\tau}_{{\tau_v=\up,\tau_{v^\prime}=\dn}})=\phi(\bm{\tau}_{\tau_v=\dn,\tau_{v^\prime}=\up})$
for any $\bm{\tau}$ when $\as{v}$ and $\as{v^\prime}$ are directly connected. 
Since $\{\as{v}\}_{v\in\calVW_L}$ is connected, we conclude that $\phi(\bm{\sigma})=\phi(\bm{\tau})$
if $\sum_{v\in\calVW_L}\sigma_v=\sum_{v\in\calVW_L}\tau_v$ 
(we regard $\up$ and $\dn$ as $+1$ and $-1$, respectively, in the sum).
Therefore, any ground state $\Phi$ of $H$ is expressed as a linear combination of $\Phi_{\up}$ 
in \eqref{eq:ferro} and its SU(2) rotations. 
\end{appendices}

\end{document}